\documentclass[sigconf]{acmart}
\usepackage{listings}
\usepackage{siunitx}
\AtBeginDocument{%
  }

\copyrightyear{2024}
\setcopyright{none}
\acmYear{2024}

\acmConference[CHCI 2024]{Make sure to enter the correct
  conference title from your rights confirmation emai}{August 19--21,
  2024}{Taiyuan, China}

\renewcommand\footnotetextcopyrightpermission[1]{} 

\begin{document}

\title{AR Secretary Agent: Real-time Memory Augmentation via LLM-powered Augmented Reality Glasses}

\author{Raphaël A. El Haddad}
\orcid{0009-0008-4925-1069}
\authornotemark[1]
\email{elhaddadr10@mails.tsinghua.edu.cn}
\email{raphael_elhaddad_edu@outlook.com}
\affiliation{%
  \institution{Tsinghua University}
  \state{Beijing}
  \country{China}}

\author{Zeyu Wang}
\email{wang-zy23@mails.tsinghua.edu.cn}
\orcid{0009-0007-5048-1665}
\affiliation{%
  \institution{Tsinghua University}
  \state{Beijing}
  \country{China}}

\author{Yeonsu Shin}
\email{srs23@mails.tsinghua.edu.cn}
\email{shin.yeonsu1007@outlook.com}
\orcid{0009-0006-6678-5228}
\affiliation{%
  \institution{Tsinghua University}
  \state{Beijing}
  \country{China}}

\author{Ranyi Liu}
\email{liu_ry01@163.com}
\orcid{0009-0007-2572-1751}
\affiliation{%
  \institution{Tsinghua University}
  \state{Beijing}
  \country{China}}

\author{Yuntao Wang}
\email{yuntaowang@tsinghua.edu.cn}
\orcid{0000-0002-4249-8893}
\affiliation{%
  \institution{Tsinghua University}
  \city{Haidian Qu}
  \state{Beijing Shi}
  \country{China}}

\author{Chun Yu}
\email{chunyu@tsinghua.edu.cn}
\orcid{0000-0003-2591-7993}
\affiliation{%
  \institution{Tsinghua University}
  \city{Haidian Qu}
  \state{Beijing Shi}
  \country{China}}

\renewcommand{\shortauthors}{El Haddad et al.}

\begin{abstract}
Interacting with a significant number of individuals on a daily basis is commonplace for many professionals, which can lead to challenges in recalling specific details: Who is this person? What did we talk about last time? The advant of augmented reality (AR) glasses, equipped with visual and auditory data capture capabilities, presents a solution. In our work, we implemented an AR Secretary Agent with advanced Large Language Models (LLMs) and Computer Vision technologies. This system could discreetly provide real-time information to the wearer, identifying who they are conversing with and summarizing previous discussions. To verify AR Secretary, we conducted a user study with 13 participants and showed that our technique can efficiently help users to memorize events by up to 20\% memory enhancement on our study.
\end{abstract}



\begin{teaserfigure}
  \includegraphics[width=\textwidth]{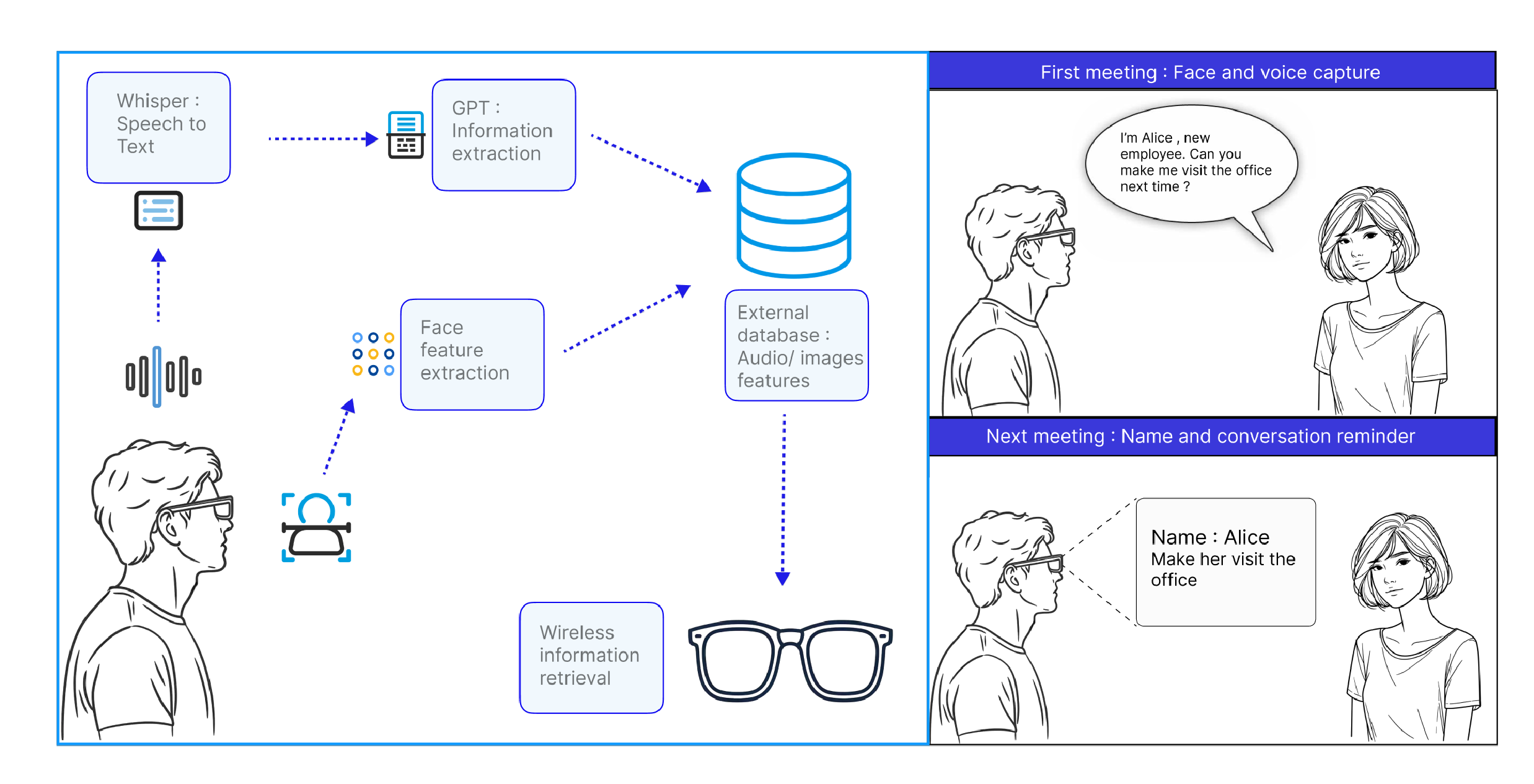}
  \caption{Secretary Assistant concept and general pipeline}
\end{teaserfigure}


\maketitle

\section{Introduction}
Memory plays a key role in people's life~\cite{baddeley2020memory, decision_making} yet it is unreliable due to forgetting~\cite{murre2015replication} and diseases~\cite{dementia}. One promising solution,
Personal Memex~\cite{bush1945we}, which refers to the capability
to capture and record personal experiences in daily life, can be considered as an external memory augmentation to human. With advancements in the computational capabilities of wearable devices, the incorporation of an always-on virtual assistant that can understand user's daily experience and provide summary on-demand has the potential to greatly enhance user's memory. Specifically, AR glasses provide a convenient solution by seamlessly displaying information to the user and becoming closer to regular glasses~\cite{light_AR_glasses}.

Lots of HCI research had been done on boosting human memory with smart devices, such as life-recording systems that continuously capture user's experience through multimedia and sensor data~\cite{memento, mediated_reality, audio_memeory_aid, chang2021memx}, as well as real-time information retrieval systems that provide contextual information~\cite{vimes, memory_glasses, pal_wearable, com_motion, just_in_time}. While these technologies are promising, they often demand the user's full attention, highlighting the need for more seamless, non-disruptive memory support. Most importantly, they lack the ability to facilitate memory of information intensive scenarios, such as presentations or meetings, which is a daily basis for many professionals. To fill-in the gap, our work aims to answer the research questions as follows:

\begin{enumerate}
    \item How to properly use AR glasses as a memory augmentation technique that can seamlessly capture and respond to users on-demand?
    \item Combining with cutting-edge algorithms, can AR glasses handle information intensive scenarios and even surpass regular note taking approaches?
\end{enumerate}

In this paper, we developed AR Secretary that enhance user's memory by capturing the user's visual field and surrounding audio events. During each conversation/meeting, our AR secretary captures the content through audio recording and automatically summarize it with Large Language Models. The summary of each voice events were saved to a database with the speaker's identity as entry. Upon recognition of a face that matches an entry in the existing database, the system displays the individual's name and a summary of the previous interaction with them on the AR glasses. This process is initiated wirelessly through a smart ring, which is paired with the glasses via Bluetooth, enabling discreet and seamless operation. 

To evaluate the efficiency and usefulness of our AR secretary, we conducted a user study with 12 participants. During the experiment, participants were involved in four conversations with dense information and afterwards, we tested their short-term memory and long-term memory. We tested our AR secretary's usage under two comparison setting: (1) how it performs comparing to traditional note taking approach, and (2) how helpful it is depending on user's note taking willingness. The result show that our technique can efficiently help users to memorize events by up to 20\% memory enhancement. We also conducted interview afterwards.

To conclude, our contribution can be summarized as follows: 
\begin{itemize}
    \item We proposed and implemented a novel AR Secretary which functions as external memory enhancement. Our AR secretary can function seamlessly and respond to users on-demand.
    \item Through a comprehensive user study, we verified the usability of our AR secretary and provided insights for future memory augmentation technique developments.
\end{itemize}

\section{RELATED WORK}
Enhancing memory through wearable devices has been an area of active research. Additionally, the capabilities of large language models (LLMs) for summarization and comprehension are rapidly advancing. However, the integration of AR glasses with LLMs for memory enhancement remains relatively unexplored.

\subsection{Wearable Memory Assistant}
Lifelogging systems, which utilize multi-modal inputs such as images, audio, biometric data, and movements to enhance daily life, have been developed since the 1990s. These devices, often termed "memory prostheses" \cite{memory_prosthesis}, aim to augment human memory. Voice-based data augmentation devices are particularly notable. In 2004, Vemuri et al. \cite{audio_based_memeory_aid} introduced a memory assistant that records audio and allows users to retrieve information via keywords. Similarly, Hayes et al. \cite{pal_audio_loop} developed the PAL (Personalized Audio Loop) device, which continuously records audio to create a "memory timeline." Yamano and Itou \cite{audio_life_log} further improved information retrieval by utilizing audio metadata, while Shah et al. \cite{life_logging_archive} post-processed audio data to identify patterns and enhance retrieval.

These devices typically require the user to interact with a phone or screen, which can detract from their seamless integration into daily activities. Gelonch et al. \cite{acceptability_lifelogging} investigated the acceptability of a wearable lifelogging camera and found that ease of use was a critical factor. Despite advancements, many of these systems rely heavily on audio data, which can be limiting due to environmental noise and microphone quality. The advent of computer vision offers promising avenues for improving memory augmentation.

\subsection{Face Detection as a Memory Assistant}
Image and video-based facial recognition technology has become highly efficient and can operate without the subject’s awareness \cite{review_face_recognition} \cite{review_face_recognition_2}. Much of the research on facial recognition for memory augmentation has focused on aiding visually impaired individuals. For instance, G. Divya et al. \cite{presence_detection} developed a presence detector that provides feedback to users. Zhao, Yuhang, and Wu et al. \cite{messenger_face_visually_impaired} created a system that helps visually impaired users recognize friends on Facebook through facial recognition.

The limited computational power of current AR glasses has hindered their widespread adoption for face detection applications. Some researchers have explored cloud or fog computing solutions to address this limitation, as demonstrated by Hellmund et al. \cite{fog_computing_glasses}, or leveraging smartphones for additional computing power, as shown by Mavridou et al. \cite{smartphone_glasses_computing}.

\subsection{Virtual Assistants and LLMs}
Voice queries are the fastest method of input on smart devices \cite{speech_fastest}, and they align closely with natural communication methods, as noted by Guy \cite{speech_most_matural}. LLMs are particularly adept at processing audio data due to their powerful comprehension and summarizing capabilities. One significant challenge in audio-based comprehension is handling hesitations or inaccuracies in speech, which LLMs can manage effectively according to Van Gysel \cite{advantage_LLM_voice_assistant}. Goyal et al. \cite{summarization_LLM} found that LLMs excel in summarizing information, making them suitable for organizing unstructured audio data for memory augmentation.

The "Memoro" project \cite{memoro} combines LLMs with wearable devices but relies on audio output instead of visual displays. As holographic technology advances \cite{holographic_technology}, AR glasses are poised to become the new standard for smart devices. Utilizing their display capabilities for user information delivery appears more future-proof, less intrusive, and more socially acceptable.

\section{DESIGN AND IMPLEMENTATION}
The application has been designed to run on the INMO AIR 2 glasses, paired with the INMO Ring 2.

\subsection{AR Glasses Interface}

\begin{figure}[tph!]
\centerline{\includegraphics[totalheight=3cm]{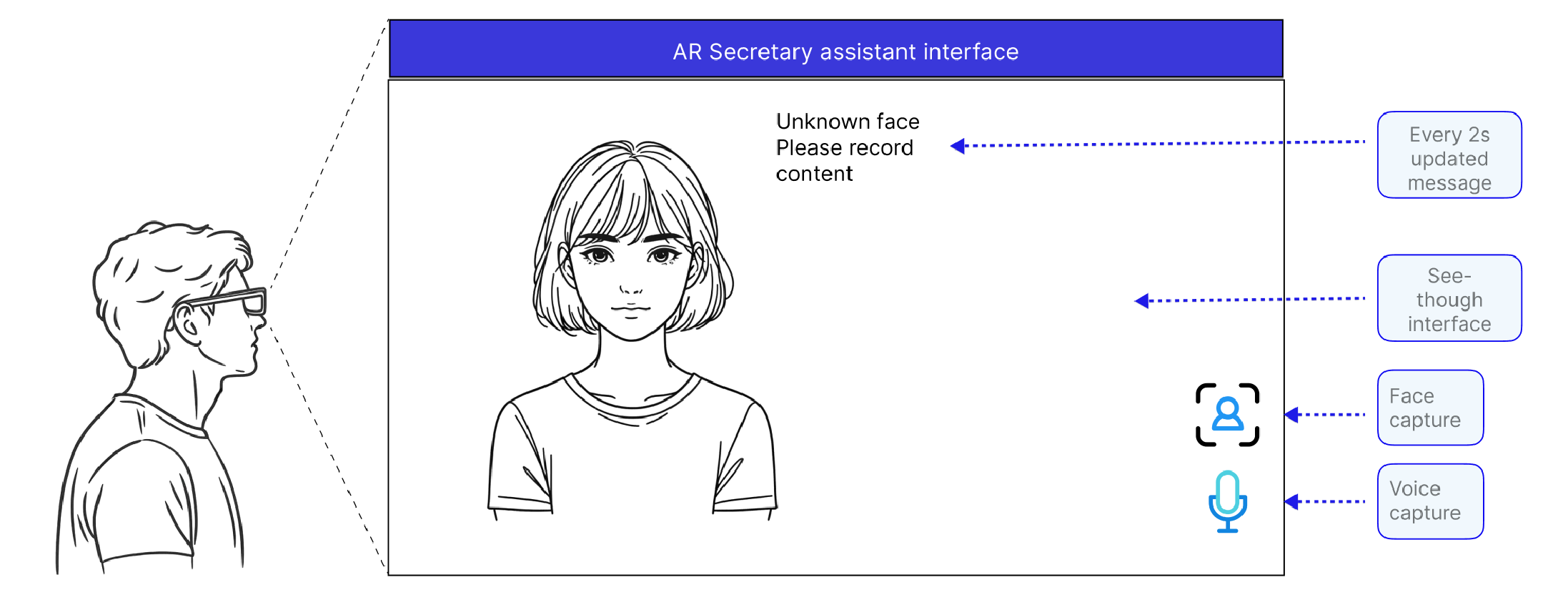}}
    \caption{Augmented Reality Interface}
    \label{fig:concept}
\end{figure}

The application was developed using Android Studio and operates on Android 9.0. The user interface (UI) is designed to be minimalist to avoid obstructing the user's field of view. It includes text display when a face is detected and recognized, along with two buttons located in the bottom-right corner: one for capturing images and one for starting and stopping audio recordings. \\
The device stores the captured data (audio or images) and transmits it to a server within the same network, using the traditional DATE-TIME format for filenames. A more comprehensive solution utilizing the World Wide Web could be considered for future implementations. \\
The glasses send requests to the server every 2 seconds to retrieve processed data.

\subsection{Audio Processing}
Upon reaching the server, the audio data is entered into an SQL database, and the processing begins. \\
The audio file is segmented into 30-second intervals and processed using Whisper \cite{whisper_openai}, which converts the audio into a full transcript. Whisper also includes noise filtering capabilities; however, further tests indicated that pre-processing noise removal does not significantly enhance results. The full transcript is stored in the SQL database. \\
This transcript is then sent to OpenAI servers via their API and processed by GPT-4. Prompt engineering has been utilized to extract the following information:
\begin{itemize}
    \item \textbf{Name}: If not explicitly stated, a role or descriptor of the person
    \item \textbf{To-do}: Tasks identified from the transcript
    \item \textbf{Summary}: A concise, one-sentence summary of the content
\end{itemize}
All extracted information is subsequently added to the database. \\
For a 30-second audio clip, the entire process from sending data from the glasses to finalizing the instance in the database takes less than 20 seconds.

\subsection{Image Processing}
When an image reaches the server, it is immediately converted into features using the Python module face-recognition. These features include the size of the ears, the spacing between the eyes, the length of the nose, etc., resulting in a 128-feature matrix per face. The comparison is performed using a lightweight linear SVM model, which executes in milliseconds. \\
\\
If a face is deemed \textbf{similar enough} to an existing entry, the corresponding data is retrieved, and the variable sent to the glasses is updated. The glasses, fetching this global variable every 2 seconds, will display the relevant information. The face encoding matrix is saved in the known faces database, and multiple face entries for the same person enhance the model’s accuracy. \\
\\
If \textbf{no known faces are similar}, the image is associated with the audio recorded immediately after the image capture, identified by the DATE-TIME filename format. A new entry is created in the image dataset, and the name/role associated with the image is extracted by GPT-4 from the subsequent audio. The feature matrix is added to the known faces database, and its position is recorded in the image dataset.

\subsection{Dealing with Bottlenecks}
While image processing is swift, increasing database size can slow down SVM processing. Initial tests demonstrated high speed, though testing with larger datasets is necessary. \\
Audio processing can also be time-consuming, particularly for longer recordings. \\
\\
To prevent new data processing delays due to server workload, tasks are offloaded using Celery, a distributed system compatible with Flask. Heavy tasks are processed on separate threads, with Celery managing the queue system.

\subsection{Multiple Meetings}
If a known face is detected, the subsequent speech recording is associated with that person. The transcript is concatenated with previous entries, along with timestamps. The summary is updated to reflect the latest conversation.

\subsection{Data Retrieval}
All data is stored in an SQL database, and a web application was developed to facilitate data retrieval. Users can view pictures and names of saved contacts and review previous conversations. This feature assists in recalling past meetings. Future enhancements could include an LLM in the web application to enable complex queries, such as: "What tasks did I discuss with my colleague yesterday?" or "How many students are working on AR glasses projects?"

\begin{figure*}[tph!]
\centering
\includegraphics[width=14cm]{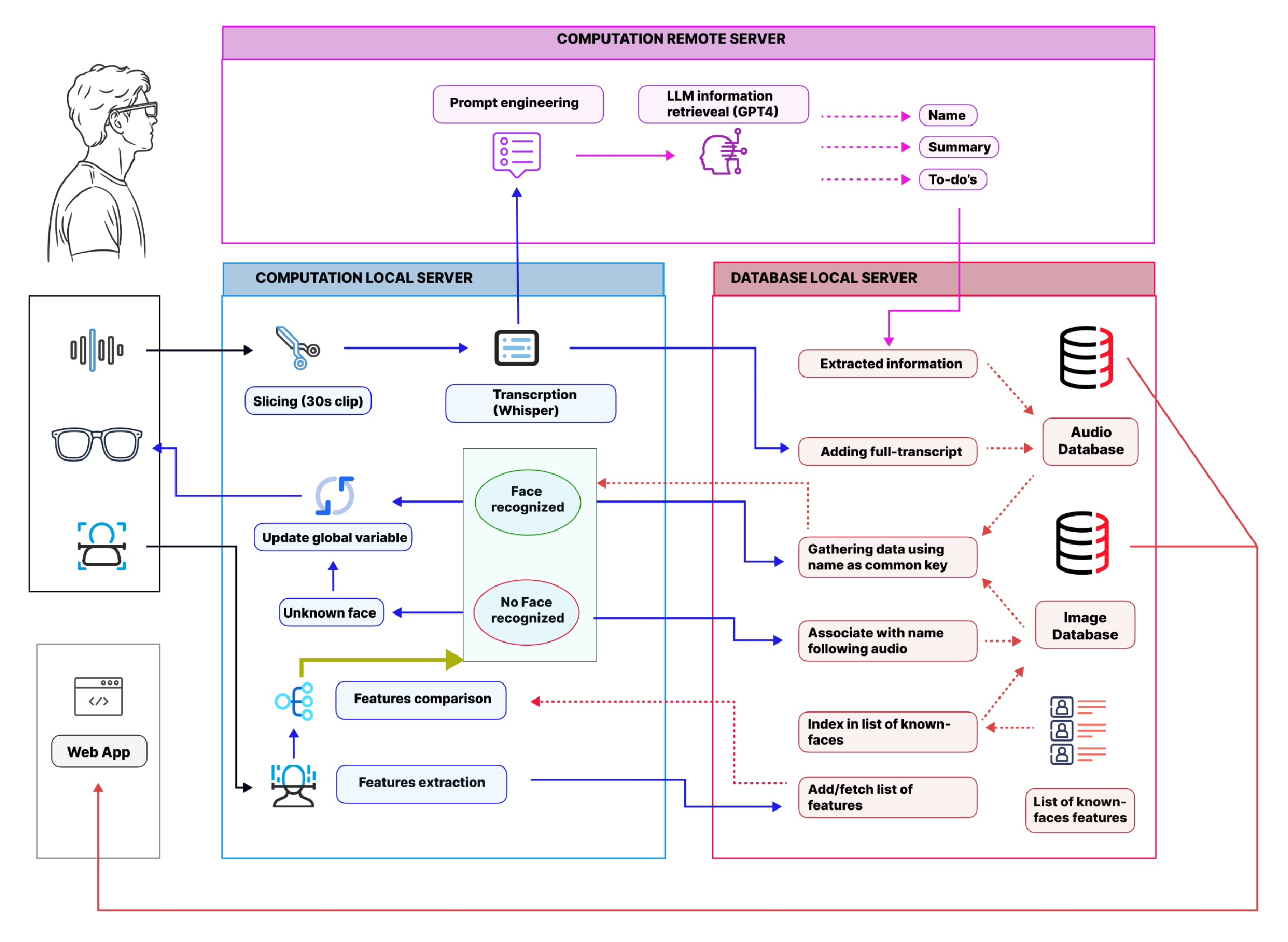}
\caption{Detailed Pipeline}
\label{fig:pipeline}
\end{figure*}

\section{USER STUDY}
To evaluate the effectiveness of our tool, we conducted a user study with 12 participants recruited from the campus. The study aimed to assess the performance of our tool for both short-term and long-term memory enhancement. Three baselines were considered for comparison: no memorization effort, memorization effort, and taking notes.

\subsection{Tasks}

\subsubsection{Short-term Memory}
\paragraph{Memory Only}
Participants were invited to listen to four speeches delivered by four different individuals, each lasting 3 minutes. One-third of the participants were unaware of the study's goal (no memorization effort), another third were informed they needed to remember as many details as possible, and the final third were allowed to take notes on their phones. \\
After the speeches, participants were asked to write down the names and content of the conversations.

\paragraph{Help of the Glasses}
Participants were then taught how to use the glasses, scan faces, and record audio. All speeches were recorded using the glasses. Long summaries (4 to 5 sentences) generated from these recordings were sent to the participants, who were asked to provide additional information about the speeches, prompted by the summaries.

\subsubsection{Long-term Memory}
\paragraph{Memory Only}
After 3 to 4 days, without prior notice, participants were asked to recall the names and content of two of the four speakers they had previously met (chosen randomly).

\paragraph{Help of the Glasses}
Subsequently, participants were sent short summaries (1 to 2 sentences) generated by the glasses, which they would see if they scanned the face of the person. They were then asked to provide additional information about the speeches, prompted by the summaries.

\subsubsection{Metrics}
The speeches were prepared with a high number of keywords, including tricky and important information. These keywords were highlighted in advance. To assess participants' memorization capabilities, the number of keywords they typed in the survey, divided by the total number of keywords, represented their \textbf{score}. The number of additional keywords found using the summary, divided by the total number of keywords, represented their \textbf{improvement score}. \\
The goal was to compare the significance of the improvement score across each baseline (with or without memory effort).

\subsubsection{Qualitative Analysis}
In addition to the quantitative analysis, participants completed a questionnaire rating their memory, the contexts in which they usually take notes, their assessment of the tool (ease of use, efficiency), and their suggestions for improvement.

\subsection{Speakers and Participants}
The four speakers were non-native English speakers. All speeches were delivered in English, using fictional English names and background stories. The speeches were conducted face-to-face in a quiet environment. \\
Twelve participants took part in the study (3 females, 9 males, aged 22 to 36, $\text{age}_{\text{mean}} = 25$, $\text{age}_{\text{std}} = 3.44$). All participants were fluent in English, with two being native speakers. Participants rated their memory abilities on a scale from 1 to 10 (range from 3 to 8, $\text{memory}_{\text{mean}} = 6.5$, $\text{memory}_{\text{std}} = 1.44$). None had significant prior experience with AR glasses. \\
Participants also shared their note-taking habits: four participants usually never take notes, two always take notes, and the remaining participants take notes for important events (e.g., significant meetings or classes).

\section{RESULTS}

\subsection{Quantitative Study: First Part}
We collected data on what participants remembered immediately after listening to the four speeches. \\
For each speaker, we calculated the percentage of the speech remembered by counting the number of keywords recalled, divided by the total number of keywords. We then calculated the improvement by counting the number of additional keywords recalled with the help of the summary.

\subsubsection{Data Analysis}
The following tables display the mean and standard deviation for different scenarios. Table \ref{tab:speaker_general} shows the figures for each speaker. Table \ref{tab:memory_mode_general} presents data for the three different types of participants: those unaware of the memorization task (No effort), those aware but not allowed to take notes (Only memory), and those allowed to take notes (Phone notes).

\begin{table*}[tph!]
\centering
\small
\resizebox{\textwidth}{!}{%
\begin{tabular}{l
                S[table-format=3.1]
                l
                S[table-format=3.1]
                l
                S[table-format=3.1]
                l
                S[table-format=3.1]
                l
                S[table-format=3.1]
                l}
    \toprule
    \textbf{Condition} & \multicolumn{2}{c}{\textbf{Macro-average}} & \multicolumn{2}{c}{\textbf{Josh}} & \multicolumn{2}{c}{\textbf{Charlotte}} & \multicolumn{2}{c}{\textbf{Sophia}} & \multicolumn{2}{c}{\textbf{Sarah}}\\
    \midrule
    With memory                     & 39.6 & ($\sigma = 17.3$) & 48.6 & ($\sigma = 20.5$) & 36.4 & ($\sigma = 16.0$) & 34.0 & ($\sigma = 19.5$) & 39.5 & ($\sigma = 23.4$)\\
    Improvement with summary      & 12.4 & ($\sigma = 8.6$) & 9.9 & ($\sigma = 10.2$) & 16.4 & ($\sigma = 13.5$) & 9.6 & ($\sigma = 6.8$) & 13.6 & ($\sigma = 10.5$)\\
    \bottomrule
\end{tabular}}
\caption{Percentage of content remembered by participants and the improvement provided by the summary for each speaker and overall. The results are for the first user study focusing on short-term memory.}
\label{tab:speaker_general}
\end{table*}

\begin{table*}[tph!]
\centering
\small
\resizebox{\textwidth}{!}{%
\begin{tabular}{l
                S[table-format=3.1]
                l
                S[table-format=3.1]
                l
                S[table-format=3.1]
                l
                S[table-format=3.1]
                l}
    \toprule
    \textbf{Condition} & \multicolumn{2}{c}{\textbf{Macro-average}} & \multicolumn{2}{c}{\textbf{No effort}} & \multicolumn{2}{c}{\textbf{Only memory}} & \multicolumn{2}{c}{\textbf{Phone notes}} \\
    \midrule
    With memory                     & 39.6 & ($\sigma = 17.3$) & 27.8 & ($\sigma = 7.3$) & 29.6 & ($\sigma = 11.4$) & 56.8 & ($\sigma = 11.7$)\\
    Improvement with summary      & 12.4 & ($\sigma = 8.6$) & 20.6 & ($\sigma = 2.4$) & 17.5 & ($\sigma = 4.0$) & 2.3 & ($\sigma = 2.4$)\\
    \bottomrule
\end{tabular}}
\caption{Percentage of content remembered by participants and the improvement provided by the summary, categorized by participants’ awareness and note-taking status: no effort, only memory, and phone notes. The results are for the first user study focusing on short-term memory.}
\label{tab:memory_mode_general}
\end{table*}

\subsubsection{Statistical Analysis}
\begin{figure}[tph!]
\centerline{\includegraphics[totalheight=4.5cm]{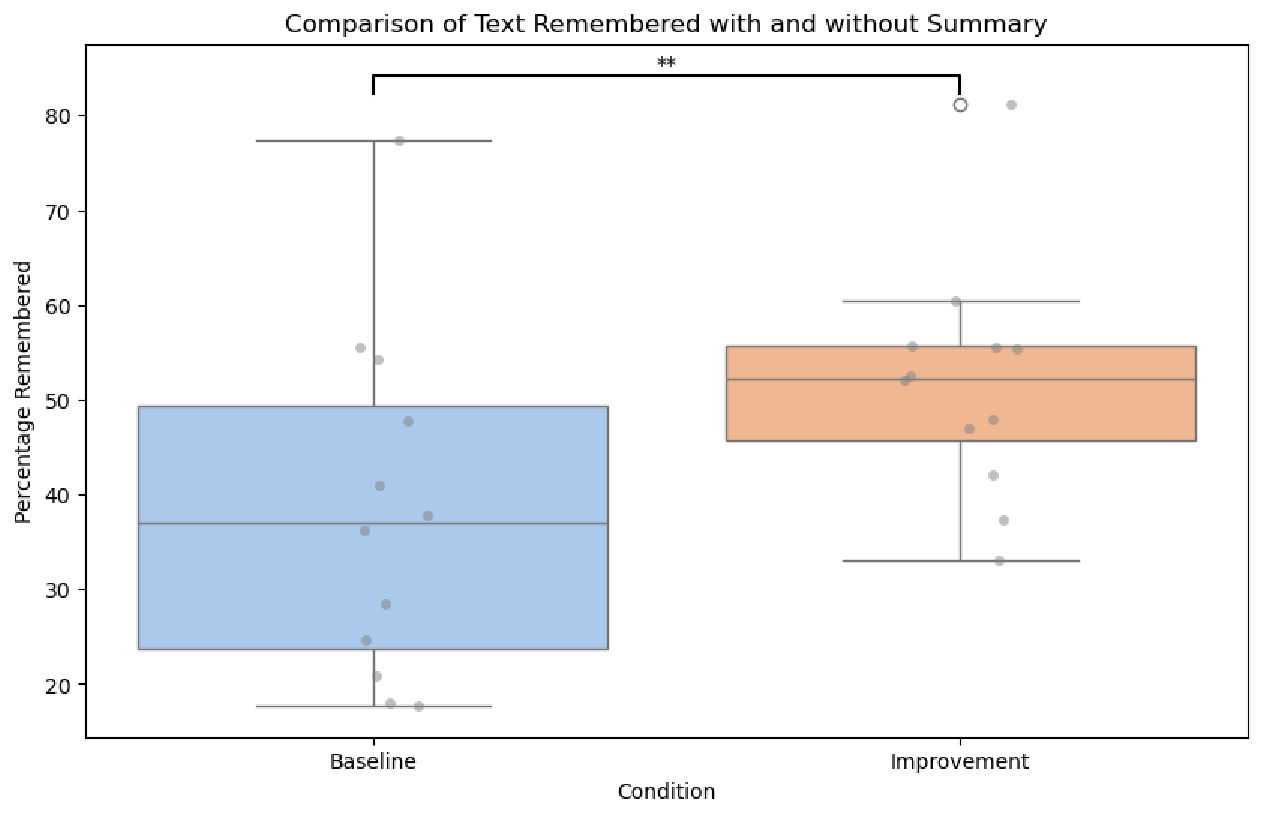}}
    \caption{Box-plot and Wilcoxon test (** $p < 0.01$) on overall averaged percentage of speech remembered, without and with help of AR assistant, for short term-memory}
    \label{fig:quantitative_general}
\end{figure}

\paragraph{Macro Analysis}
The first Null Hypothesis considered is that the "AR Secretary Agent does not help to retrieve more information." To test this, we averaged the results across the four speakers, without initially considering the different types of participants (those who can take notes or not). \\
We conducted a Wilcoxon test (Figure \ref{fig:quantitative_general}) ($T_{\text{stat}} = 3.9$, $p_{\text{value}} = 6.1e-03$), which is suitable for non-parametric distributions and analyses with few paired data points. As all participants performed both experiments (without summary and then with summary), a paired-data analysis was appropriate. The test rejected the null hypothesis.

\begin{figure}[tph!]
\centerline{\includegraphics[totalheight=4.5cm]{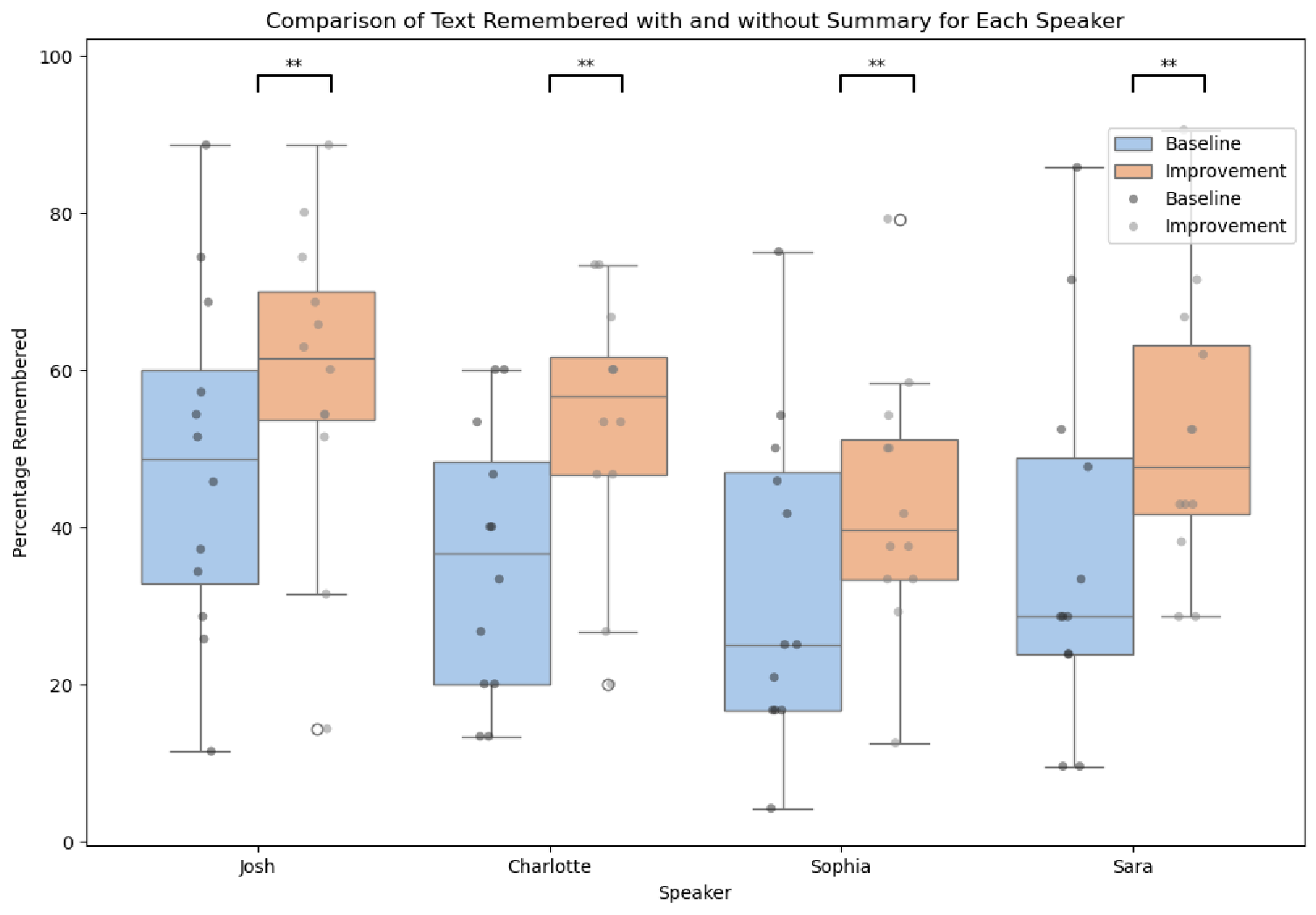}}
    \caption{Box-plot and Wilcoxon test (*** $p < 0.001$) on averaged percentage of speech remembered, for every individual speaker, without and with help of AR assistant, for short term-memory}
    \label{fig:quantitative_speaker}
\end{figure}

\paragraph{Micro Analysis}
Following the same protocol, we conducted a statistical analysis comparing the task completion for each speaker individually. The Wilcoxon tests showed significant improvements for all speakers: Josh ($T_{\text{stat}} = 0.0$, $p_{\text{value}} = 7.6e-03$), Charlotte ($T_{\text{stat}} = 0.0$, $p_{\text{value}} = 5.0e-03$), Sophia ($T_{\text{stat}} = 0.0$, $p_{\text{value}} = 4.5e-03$), and Sarah ($T_{\text{stat}} = 0.0$, $p_{\text{value}} = 5.0e-03$). The tests are summarized in Figure \ref{fig:quantitative_speaker}.

\begin{figure}[tph!]
\centerline{\includegraphics[totalheight=4.5cm]{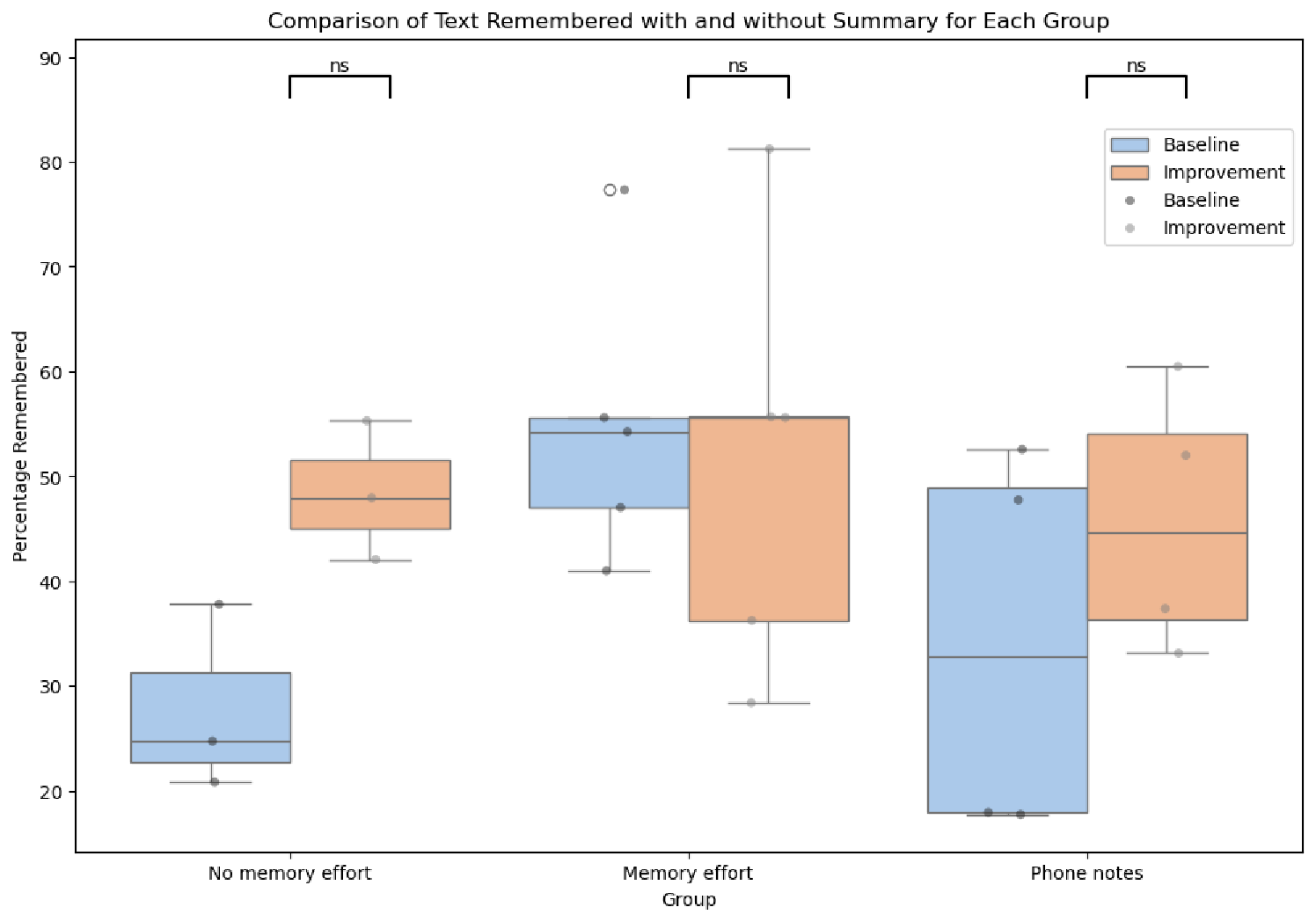}}
    \caption{Box-plot and Wilcoxon test (ns $p > 0.05$) on overall averaged percentage of speech remembered, for every memory baseline, without and with help of AR assistant, for short term-memory}
    \label{fig:quantitative_baselines}
\end{figure}

\paragraph{Comparing Baselines}
We examined three sub-baselines: participants unaware of the need to memorize the speeches (No effort), those aware but not allowed to use notes (Only memory), and those allowed to take notes (Phone notes). Due to the small sample size, we only conducted Wilcoxon tests for these sub-groups (Figure \ref{fig:quantitative_baselines}): no memory ($T_{\text{stat}} = 0.0$, $p_{\text{value}} = 0.25$), memory effort ($T_{\text{stat}} = 6.0$, $p_{\text{value}} = 0.81$), and phone notes ($T_{\text{stat}} = 2.0$, $p_{\text{value}} = 0.38$). None of these tests rejected the null hypothesis.

\paragraph{Name Remembering}
We also tested the ability of participants to remember the correct names of the speakers, which is crucial for practical use. Participants wrote down the names, and in some cases, the AR Secretary Agent displayed the names. Table \ref{tab:name_remember_general} summarizes these results.

\begin{table}[tph!]
\centering
\small
\begin{tabular}{l
                S[table-format=3.1]
                l}
    \toprule
    & \multicolumn{2}{c}{\textbf{Average + std}}  \\
    \midrule
    \textbf{Only memory} & 2.5 & ($\sigma = 1.50$) \\
    \textbf{Summary if name forgotten} & 0.5 & ($\sigma = 0.63$) \\
    \bottomrule
\end{tabular}
\caption{Average number of names remembered by participants and additional names recalled with the help of the summary if the name was initially forgotten.}
\label{tab:name_remember_general}
\end{table}

\begin{figure}[tph!]
\centerline{\includegraphics[totalheight=4.5cm]{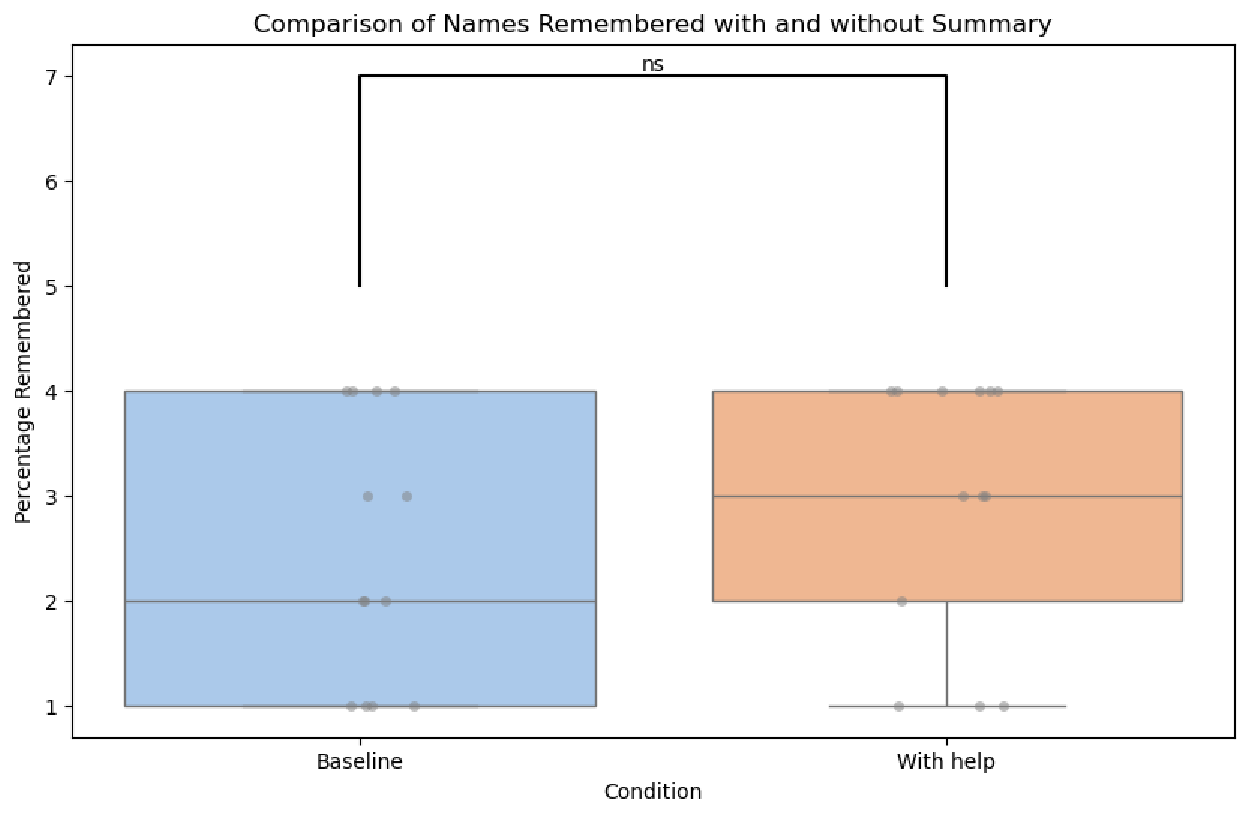}}
    \caption{Box-plot and Wilcoxon test (ns $p > 0.05$) on averaged number of names remembered, without and with help of AR assistant, for short term-memory}
    \label{fig:quantitative_name_general}
\end{figure}

To assess the effectiveness of the application in name recalling, we conducted tests on the data (names remembered without the tool and the total number of names remembered with memory plus the help of the AR Secretary Agent): a Wilcoxon test (Figure \ref{fig:quantitative_name_general}) ($T_{\text{stat}} = 9.5$, $p_{\text{value}} = 0.43$) and a T-test ($T_{\text{stat}} = -0.93$, $p_{\text{value}} = 0.36$). Although the median number of names recalled was higher with the summary (3 vs. 2), neither test rejected the null hypothesis.

\subsection{Quantitative Study: Second Part}

\subsubsection{Data Analysis}

Three days after the initial study, participants were asked to recall the names and content of two randomly selected speakers from among Josh, Sarah, and Sophia. Charlotte was excluded as the glasses did not provide an adequate summary for her speech. After collecting the participants' recollections without any aid, we provided them with the summaries generated by the glasses, allowing them to supplement their recollections with information triggered by the summaries. Table \ref{tab:name_remember_general} reports the general data, showing the average and standard deviation for each speaker and the macro-average.

\begin{table*}[tph!]
\centering
\small
\resizebox{\textwidth}{!}{%
\begin{tabular}{l
                S[table-format=3.3]
                l
                S[table-format=3.3]
                l
                S[table-format=3.3]
                l
                S[table-format=3.3]
                l}
    \toprule
    \textbf{Condition} & \multicolumn{2}{c}{\textbf{Macro-average}} & \multicolumn{2}{c}{\textbf{Josh}} & \multicolumn{2}{c}{\textbf{Sophia}} & \multicolumn{2}{c}{\textbf{Sarah}} \\
    \midrule
    No help & 17.5 & ($\sigma = 11.2$)       & 33.3 & ($\sigma = 15.5$)       & 9.4 & ($\sigma = 8.0$) & 9.8 & ($\sigma = 13.5$) \\
    Improvement with help & 14.0 & ($\sigma = 3.8$)        & 8.9 & ($\sigma = 7.3$)         & 17.8 & ($\sigma = 5.4$) &         15.3 & ($\sigma = 8.0$) \\
    Name retrieved no help & 57.7 & ($\sigma = 49.4$)        & 100.0 & ($\sigma = 0.0$)        & 50.0 & ($\sigma = 50.0$) &          22.2 & ($\sigma = 41.6$) \\
    \bottomrule
\end{tabular}}
\caption{Comparison of the average percentage and standard deviation of content remembered without help, improvement with the help of the glasses summary, and names retrieved without help, for each speaker, in the second user study focusing on long-term memory.}
\label{tab:name_remember_general}
\end{table*}

\subsubsection{Statistical Analysis}

\paragraph{Speech Content}
We conducted a macro-statistical analysis by averaging the results across all speakers. We compared the percentage of speech content participants remembered after three days without any external help to the total percentage remembered with the help of the glasses summary. A Wilcoxon test (Figure \ref{fig:post_study_general_content}) was performed ($T_{\text{stat}} = 0.0$, $p_{\text{value}} = 1.2e-05$) and the Cohen's d value was computed ($d = -0.89$), indicating a large effect ($|d| > 0.8$).

\begin{figure}[tph!]
\centerline{\includegraphics[totalheight=4.5cm]{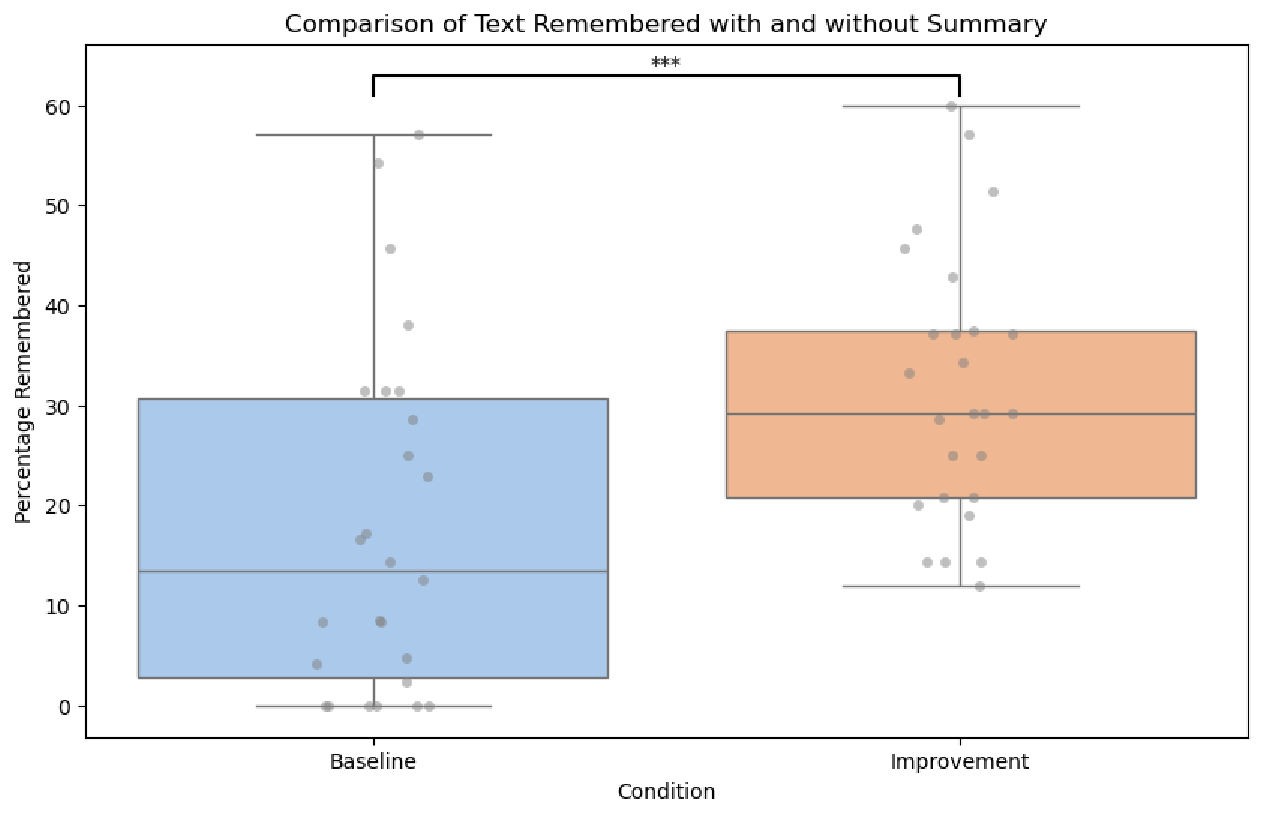}}
    \caption{ Box-plot and Wilcoxon test (*** $p < 0.001$)on overall averaged percentage of speech remembered, without and with help of AR assistant, for long term-memory}
    \label{fig:post_study_general_content}
\end{figure}

We then conducted the same analysis for each speaker individually, yielding the following results: for Josh, a Wilcoxon test ($T_{\text{stat}} = 0.0$, $p_{\text{value}} = 1.2e-02$) and Cohen’s d value ($d = -0.61$, medium effect); for Sarah, a Wilcoxon test ($T_{\text{stat}} = 0.0$, $p_{\text{value}} = 3.9e-03$) and Cohen’s d value ($d = -1.10$, large effect); and for Sophia, a Wilcoxon test ($T_{\text{stat}} = 0.0$, $p_{\text{value}} = 1.2e-02$) and Cohen’s d value ($d = -0.61$, medium effect). All tests indicated significant improvement, rejecting the null hypothesis.

\begin{figure}[tph!]
\centerline{\includegraphics[totalheight=4.5cm]{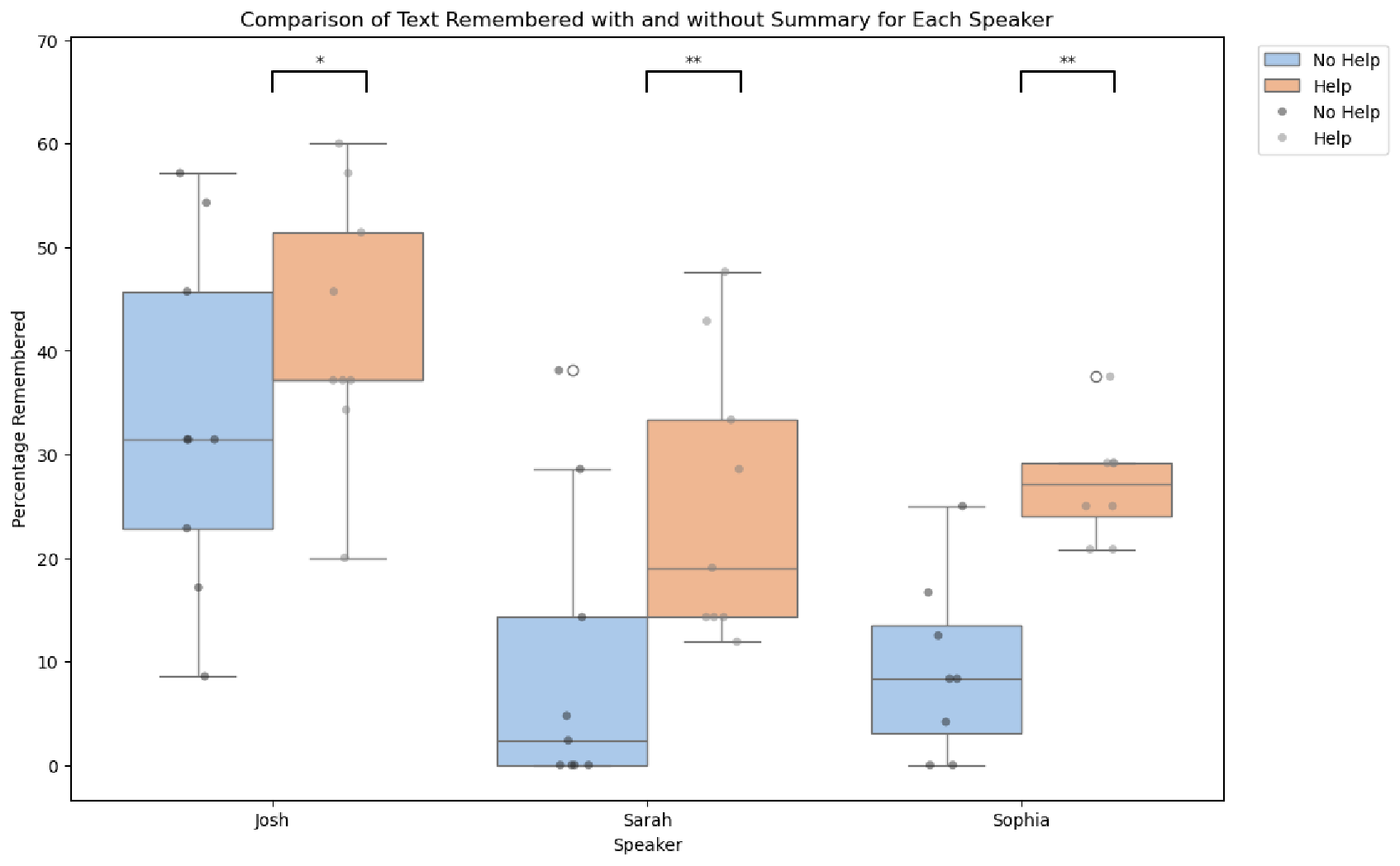}}
    \caption{Box-plot and Wilcoxon test for post-study, showing the percentage of content remembered without help and with the help of the summary for each speaker. Significant improvements were found (* $p < 0.05$, ** $p < 0.01$, *** $p < 0.001$).
     Box-plot and Wilcoxon test (* $p < 0.05$, ** $p < 0.01$, *** $p < 0.001$) on averaged percentage of speech remembered, for each individual speaker, without and with help of AR assistant, for long term-memory}
    \label{fig:post_study_general_content}
\end{figure}

\paragraph{Name Remembering}
Participants were also asked to recall the names of the different speakers. The AR glasses accurately captured the names of Josh, Sophia, and Sarah 100\% of the time. We provided a binary data list (0 if the name was not remembered, 1 if it was) and conducted a macro analysis using McNemar’s Test, averaging across all speakers ($T_{\text{stat}} = 0.0$, $p_{\text{value}} = 9.8e-04$), indicating significant help. We then performed the same test for each speaker: for Josh (not relevant as all participants remembered his name), for Sarah ($T_{\text{stat}} = 0.0$, $p_{\text{value}} = 1.6e-02$), and for Sophia ($T_{\text{stat}} = 0.0$, $p_{\text{value}} = 1.3e-01$). Significant help was observed only for Sarah.

\subsection{Qualitative Study}
We asked participants qualitative questions regarding the concept of our project, using a 5-point Likert Scale to answer seven questions.

\begin{figure}[tph!]
\centerline{\includegraphics[totalheight=4.5cm]{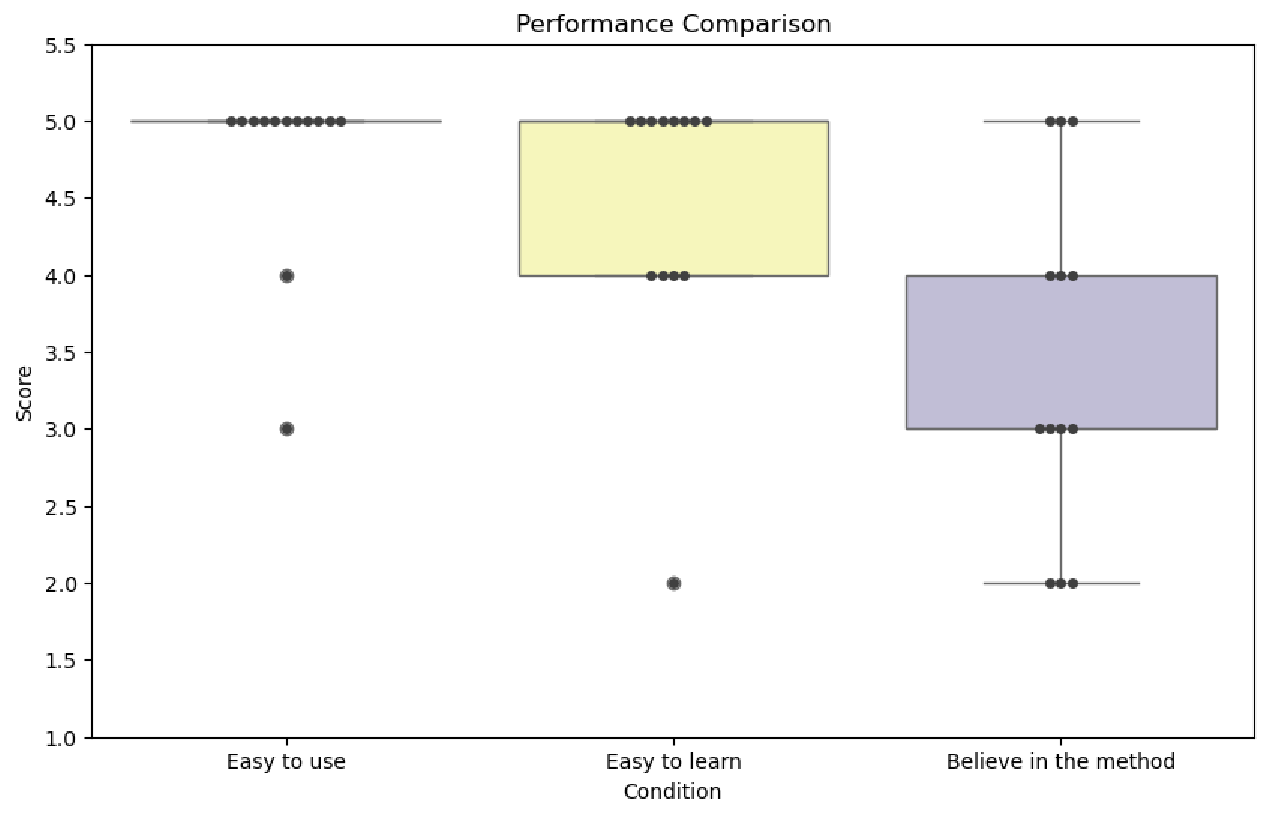}}
    \caption{Box-plot of general qualitative questions using a Likert Scale (High is better)}
    \label{fig:qualitative_generale}
\end{figure}

As shown in Figure \ref{fig:qualitative_generale}, overall satisfaction with the tool's usage was high. Ten out of twelve participants gave the maximum score for ease of use. The tool was also considered very easy to learn, with most participants mastering it within five minutes, even those with no prior experience with AR glasses. Participants generally believed in the method, although three rated their belief quite low (2/5).

\begin{figure}[tph!]
\centerline{\includegraphics[totalheight=4.5cm]{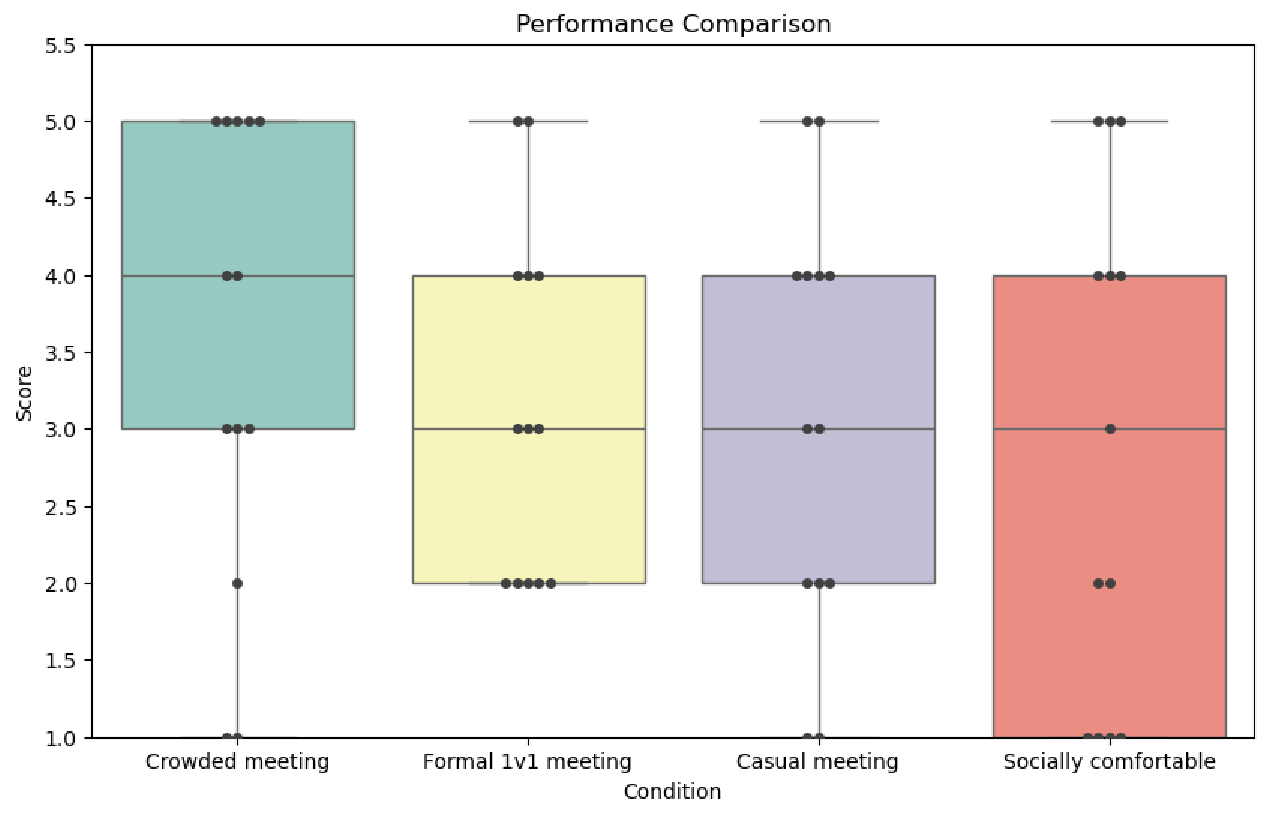}}
    \caption{Box-plot of participants' willingness to use the tool in different meeting scenarios using a Likert Scale. (High is better)}
    \label{fig:qualitative_meeting}
\end{figure}

The second box-plot (Figure \ref{fig:qualitative_meeting}) shows strong acceptance of the tool for meetings with multiple people (median = 4, mean = 4), but lower acceptance for formal meetings and casual meetings with family or friends. More than half of the participants (7/12) rated the potential use of the tool in formal meetings as 3/5 or lower. For casual meetings, the mean and median were both 3/5, indicating mixed feelings about using the tool in one-on-one interactions. \\
The last question concerned the acceptance of being recorded by others using this tool if it becomes common in daily life. As shown in Figure \ref{fig:qualitative_meeting}, the mean value (3.2) and the wide distribution indicate a division in opinions. Further insights are provided by participants' comments.

\section{DISCUSSION}

\subsection{Comments from the Participants}
At the end of the survey, participants were invited to provide their feedback and suggestions about the tool and the concept. This feedback is valuable for understanding the results of the quantitative and qualitative analyses. \\
Participants generally found the concept potentially very useful in specific cases, such as for doctors needing to remember many faces during consultations. The tool was also praised for helping recall names, which can be challenging for some people. \\
Participants suggested several improvements, such as developing a mobile app for information retrieval or controlling the recording remotely. They also recommended implementing automatic face detection and capture, and displaying a preview of the captured picture directly on the glasses. \\
Many comments focused on potential improvements. The most frequent concern was the quality of the summaries, which sometimes contained incorrect information. Participants understood that this limitation was mainly due to the microphone's quality and its ability to capture remote sounds. They also advised using lighter weight glasses. \\
Overall, the concept was praised, though still limited by current hardware capabilities.

\subsection{Short-term Memory Enhancement}
The tool enhances short-term memory, but comparisons with traditional memory enhancement methods are less favorable. The first study tested participants' short-term memory. As summarized in Table \ref{tab:speaker_general}, using the glasses resulted in a 10\% improvement in recalling conversation content. The tool also helped retrieve names (Figure \ref{fig:quantitative_name_general}), with participants recalling a median of three names versus two without the tool. Statistical analysis showed that the null hypothesis could be rejected for the overall case (Figure \ref{fig:quantitative_general}) and for each speaker individually (Figure \ref{fig:quantitative_speaker}). \\
However, baseline comparisons could be improved. The glasses yielded satisfactory results, with up to a 17\% memory increase for participants not taking notes (Table \ref{tab:memory_mode_general}), but the sample size was small, and results could be influenced by randomness. None of the baseline statistical analyses passed the Wilcoxon test successfully (Figure \ref{fig:quantitative_baselines}).

\subsection{Long-term Memory Enhancement}
The tool's usefulness for long-term memory is promising and could represent its real benefit. \\
In the second study (Table \ref{tab:name_remember_general}), the average memory enhancement was higher than in the first study (14.0\% vs. 12.4\%). Additionally, the tool was very effective in capturing names if the person was properly introduced. In cases where remembering a name was difficult (e.g., only 22\% remembered Sarah's name), the tool proved beneficial. The Wilcoxon test passed successfully for all tests, both macro and for each speaker.

\section{LIMITATIONS AND FUTURE WORK}

\subsection{Microphone Limitation}
The glasses' microphone is designed to isolate the user's voice from surrounding noise, whereas our project required capturing surrounding voices. A second microphone designed to capture ambient sounds would be a significant improvement, or alternatively, using a smartphone for sound recording as a temporary solution. \\
Experimenting with software voice enhancement as a pre-processing step to remove noise and enhance voice quality could also be beneficial.

\subsection{Multilingual Capabilities}
The current solution is set to English only, leading to incorrect transcription of non-English names or terms. Using a multilingual transcription method could address this issue, although such models still struggle with detecting words in different languages within the same context. A unilingual tool would perform better for now.

\subsection{UI Improvements}
Participants suggested several UI improvements for the glasses. First, the action to trigger the camera seems unnecessary; developing a system that automatically detects faces and takes pictures would be more user-friendly. \\
Additionally, providing more control over the UI, such as a preview of the captured image and the ability to stop recording using a smartphone, would enhance user experience.

\subsection{Multiple Voice Detection}
The tool currently does not function well in scenarios with multiple people speaking. It cannot differentiate between speakers and provides a summary of the general conversation, as if only one person is speaking. This limits its usefulness for summarizing conversations. \\
As each person has unique voice features, an additional pre-processing step could involve splitting the audio to isolate one person's voice and mute others, followed by the usual summarizing process. Although this process could be computationally intensive, it can be managed since all computations are done remotely, and only face recognition needs to be quick. Voice processing can take more time since the summary is needed only a few hours later.

\subsection{Editing Features}
Even with all software processes completed, there could be mistakes or additional information the user wants to add. Users might also wish to personalize the summary displayed for each person in the database. Adding an editing tool on the webpage that automatically updates the database would be a valuable feature.

\section{CONCLUSION}
We implemented and studied a memory augmentation system using AR glasses, powered by LLM. By adding a seamless method to record and summarize conversations, our tool manage to bring significant improvement to the memory of the user. By letting the user have access to a complete summary of its past conversation, processed by the latest version of GPT4, we demonstrated how much can the short-term memory be enhanced. 
\\
By taking advantages of the Augmented Reality glasses and by displaying the name and a quick summary of previously encountered people, we demonstrated even more promising results, especially against the bare long-term memory, for information retrieval.

\clearpage

\begin{acks}
To Yuntao Wang for supervising the project, the Human-Computer Interaction Lab for providing the hardware and the funds for the user study, to the Tsinghua classmates for their precious data for the user study.
\end{acks}

\bibliographystyle{ACM-Reference-Format}
\bibliography{sample-base}

\clearpage

\appendix

\section{SPEECHES}

\subsection{Speech 1}
My name is \textbf{Josh}. I would like to talk to you about my start-up.
I am an alumnus of \textbf{Tsinghua University}. I majored in \textbf{Computer science} and graduated recently.
I am planning to launch a company in \textbf{Chengdu} in\textbf{2025}. This start up would be about a \textbf{Smart-Ring} mixed with wedding ring. The name of the company is "\textbf{Voilier}", a fancy name using French word.
Smart rings are too bulky and wedding ring will replace it as soon as people get married. We want to develop a smart ring with \textbf{luxury aspect}, very thin (at least as thin and regular wedding ring).
The ring will have many features like \textbf{sleep tracking}, \textbf{sport} mode to analyze the calories you burnt, \textbf{heart rate}, \textbf{oxygen} in the blood.
But you can also use the ring to \textbf{control} your laptop and computer. The ring has captors, and you can do movement with the ring. You can customize the movements: drawing a shape, drawing a letter, or just swiping.
You could for example activate the camera, take a \textbf{selfie} without touching your phone, go to the \textbf{next slide} in your PPT, change the \textbf{volume} of your music by acting like a DJ or change the music.
We are working with jewel expert from \textbf{Switzerland} to build a luxury ring. We planned to release 3 models: a \textbf{diamond} one (more than 20k \$), a \textbf{gold} one (around 10k \$) and a \textbf{silver} one (3k \$).
I am currently recruiting Tsinghua Student for several positions: \textbf{Software engineer}, \textbf{marketing position}, \textbf{international business associate}. The salary is attractive, and you can have up to \textbf{6 weeks} of holiday per week.
Anyone can be allowed, whatever your \textbf{level of Chinese} is. We nevertheless require \textbf{B2 level of English}. For any business-related job, we require \textbf{C1} as we have many stakeholders in \textbf{United Kingdom} and \textbf{Ireland}.
If you have any friend that are allowed to work in Europe, we would also like to create an office in \textbf{Geneva}, to be close the expert. The salary is \textbf{twice higher}, but you’ll need to work on the \textbf{Chinese shift}, around 6h30 time gap. But the \textbf{food is free} and delicious.
I don’t have \textbf{WeChat}. I only use email and \textbf{Twitter}. I will give you my email later. As data are \textbf{confidential}, I need to encrypt my mail. I hope it is fine for you.

\subsection{Speech 2}

My name is \textbf{Sophia}. I would like to share details about my innovative project. I graduated from \textbf{MIT} with a major in \textbf{Electrical Engineering} and completed my studies recently. I am planning to establish a company in \textbf{Seoul in 2025}. This start-up will focus on \textbf{VR Glasses} designed for \textbf{educational purposes}. The company will be called "\textbf{Envision}," inspired by the vision of a smarter future.
Current VR glasses are often cumbersome and lack versatility. Our aim is to develop VR glasses that are not only technologically advanced but also sleek and lightweight, making them suitable for daily use. The glasses will offer a wide range of features, including \textbf{interactive learning modules}, \textbf{real-time translation}, and \textbf{augmented reality} enhancements.
In addition, the VR glasses will be capable of controlling various devices. They will include sensors that detect eye movements and gestures, allowing users to perform actions like \textbf{navigating educational content}, \textbf{activating virtual tutors}, or \textbf{accessing online resources} with ease. Users can personalize these interactions by setting up specific \textbf{gestures}, such as \textbf{eye blinks}, \textbf{head tilts}, or \textbf{hand motions}.
We are partnering with \textbf{top tech firms} to integrate advanced AI algorithms into our VR glasses. We plan to introduce three distinct models: one optimized for \textbf{STEM education} with built-in lab simulations, another for \textbf{language learning} with real-time translation and conversation practice, and a third model for \textbf{creative arts} featuring immersive design and drawing tools.
Additionally, we intend to establish an office in \textbf{Busan} to collaborate closely with our manufacturing partners. The compensation there is higher, but employees will need to adapt to the local work culture, including a \textbf{fourteen-hour} time difference from the United States. We offer \textbf{complimentary meals} and \textbf{accommodation}.

\subsection{Speech 3}
Hey,
Thanks so much for meeting with me today. I'm \textbf{Sarah}, an \textbf{autonomous driving algorithm engineer} with a passion for advancing mobility technology. I've been super excited about your company's work and would love to share why I'm so interested in \textbf{joining you}.
I graduated from \textbf{MIT} with a degree in \textbf{Electrical Engineering and Computer Science}, where I specialized in \textbf{machine learning} and \textbf{computer vision}. During my time there, I conducted research on autonomous vehicle algorithms, which sparked my passion for this field.
For the past few years, I've been all about making cars smarter while working with top-tier companies like \textbf{Tesla} and \textbf{Waymo}. I’ve tackled everything from \textbf{sensor fusion} to \textbf{computer vision}, teaching cars how to see, interpret their surroundings and make decisions. Did you know that most car accidents happen because of human error? I'm trying to fix that by \textbf{creating systems that can handle all sorts of driving situations safely}. One project I really enjoyed was improving our \textbf{object detection systems} in \textbf{low light conditions}. We managed to make it \textbf{20\% better} at spotting obstacles, which significantly enhances \textbf{night-time driving safety}. This is key because people need to trust that these systems work reliably, regardless of the time of day.
In addition to my professional experience, I’ve also \textbf{published papers} on autonomous driving safety protocols and presented at \textbf{industry conferences}. I believe that continuous learning and sharing knowledge are crucial in this rapidly evolving field.
What I love about your company is how you're \textbf{pioneering these innovations} and \textbf{setting industry standards}. I’m eager to bring my experience and passion to your team and contribute to your mission.
I see huge potential for us to accomplish great things together. How about we chat more over coffee next \textbf{Wednesday at 2 PM}? Before that, I will send some \textbf{materials introducing my experience to your email}. I’d love to explore how my experience aligns with the exciting projects you’re working on. Thank you once again for considering my interest. I look forward to the opportunity to contribute to your company's pioneering efforts.

\subsection{Speech 4}
My name is \textbf{Charlotte}, and I am delighted to share a little about myself with you today.
I'm \textbf{25} years old, studying \textbf{Botany}. It's a journey that continually expands my understanding of the world and our place within it.
When I'm not buried in books, you can find me \textbf{playing basketball, painting landscapes, collecting vintage stamps, or practicing the violin}. Each of these activities offers me a unique form of escape and allows me to express myself in ways words often cannot.
As for my likes, I absolutely adore \textbf{Spicy Food}. There's an interesting story behind this. My love for Spicy Food started during a trip to \textbf{Mexico}. I was dared to try a local dish known as '\textbf{Chiles en Nogada},' notorious for its fiery heat. Not only did I finish the dish, but I also found myself craving more. That experience opened up a whole new world of flavors for me, and ever since, I've been on a quest to find the perfect \textbf{balance of heat and taste}.
On the flip side, I have a strong dislike for \textbf{Horror Movies}. Let me tell you why. My dislike for horror movies can be traced back to a \textbf{sleepover at a friend's house} where we watched a horror movie marathon. I hardly slept that night and the eerie feeling lingered for days. I decided that I much prefer films that leave me inspired or in a good mood, rather than checking over my shoulder as I walk home at night.
Recently, I've been diving deep into \textbf{The Role of Photosynthesis}. It is a remarkable process plants use to convert sunlight into energy. To understand this, we compared it to \textbf{solar panels}. Just as panels convert sunlight to electricity, plants capture sunlight to produce sugars they use for food. This process not only sustains the plant but also \textbf{produces oxygen as a byproduct}, which is essential for life on Earth. Without photosynthesis, there would be no food for us to eat or air for us to breathe.

\section{QUESTIONNAIRES}

\subsection{Introduction}
Name
\\Age
\\Department in Tsinghua University
\\"How good is your memory from 1 to 10"
\\"Do you usually take notes ?"
\\"When do you take notes ?" 

\subsection{Quantitative analysis}

\begin{enumerate}
    \item \textbf{Name of the person [1, 2, 3 and 4]} : Using only memory or notes (for those allowed)
    \item \textbf{Detailed content of the presentation (no need full sentences, keywords or bag of words)} : Using only memory or notes (for those allowed)
    \item \textbf{Additional info with transcript: person [1, 2, 3 and 4]} : Using the summary given by the glasses

\end{enumerate}

\subsection{Qualitative analysis}
We measured seven aspects using a 5-points Likert Scale :

\begin{enumerate}
    \item \textbf{The tool is easy to use }
    \item \textbf{It is easy to learn how to use}
    \item \textbf{You could use this tool in a real meeting (with many people)}
    \item \textbf{You could use this tool in a formal 1vs1 meeting (with supervisor)}
    \item \textbf{You could use this tool in a casual 1vs1 meeting (friends, parents)}
    \item \textbf{Do you believe in improving memory using AR glasses?}
    \item \textbf{Would you be comfortable the person you talking to record you using glasses?}

\end{enumerate}

\subsection{Suggestions}
"The project is currently at a very early stage : Any improvements ideas ?"

\section{EXAMPLES OF SUMMARY}
The summaries could contain false information as we discussed in the limitation part.

\subsection{Speech 1}
\begin{enumerate}
    \item \textbf{Long summary}: Josh, a former student of Conquan University, wants to create a startup. The startup is about creating a smart ring that looks like a wedding ring. The ring has features like tracking, sports, calorie count, heart rate, and oxygen block. The company name would be Walee, a French name. Josh is looking to recruit a team of students for positions like software engineer, marketing, and business application. The company works with specialists in Switzerland and plans to release three models of the ring: one with diamonds, one with gold, one with silver. The company is planning to expand in Europe, specifically Geneva, Switzerland. The salary in Geneva is higher than in Beijing and food is provided for free. English proficiency at C1 level is required as they work with the United Kingdom and Ireland.",

    \item \textbf{Short summary}: Josh from the startup Wally discussed their plans for a luxury smart screen with health tracking features. They are recruiting, require English proficiency, and are collaborating with the UK and Ireland. They plan to open an office in Geneva and launch in Chengdu in 2025.
    \item \textbf{To-do's} : Email Josh if interested in the positions available, Research more about the startup Wally

\end{enumerate}

\subsection{Speech 2}
\begin{enumerate}
    \item \textbf{Long summary} : Sophia is planning to establish a company's goal in 0.0 to 5. The focus of the company is on VR glasses with a wide range of features including interactive learning modules, real-time translation, augmented reality and hand tools. The glasses will also be capable of controlling various devices. They aim to develop glasses that are not only technologically advanced but also light and flexible, making them suitable for daily use. Users can navigate educational content, activate virtual tutors, or access online resources with ease. They plan to establish an office in Wuhan to collaborate closely with manufacturing partners. They are partnering with top-tier funds to integrate advanced AI algorithms into the VR glasses.
    \item \textbf{Short summary} : Sophia discussed his innovative project of developing VR glasses with advanced features like interactive learning modules, real-time translation, and augmented reality. The glasses will also be capable of controlling various devices. They plan to establish an office in Wuhan and partner with top-tier funds for AI integration.
    \item \textbf{To-do's} : Establish an office in Wuhan, Partner with top-tier funds for AI integration

\end{enumerate}

\subsection{Speech 3}
\begin{enumerate}
    \item \textbf{Long summary} : The speaker is interested in the project and wants to contribute The speaker has experience in autonomous driving safety protocols and has published papers on it The speaker has also published a series of books The speaker has worked with companies like Tesla and Waymo The speaker has conducted research on autonomous vehicle algorithms The speaker graduated from MIT Research Green in Mexico, specialized in machine learning and computer research The speaker has improved object detection systems in low light conditions by 20\% The speaker will send some viewers introducing her experience in the next video The speaker suggests a meeting next Wednesday at 2pm"
    \item \textbf{Short summary} : The speaker, Sarah, is interested in the project and has extensive experience in autonomous driving safety protocols. She has worked with companies like Tesla and Waymo and has made significant improvements in object detection systems.
    \item \textbf{To-do's} : Prepare for the meeting next Wednesday at 2pm

\end{enumerate}

\subsection{Speech 4}
\begin{itemize}
    \item \textbf{Long summary}: Charlotte is a 25 y.o student of Botany who enjoys expressing herself through various activities such as playing basketball, painting landscapes, collecting vintage stamps, and practicing the violin. She has a passion for spicy food, developed during a trip to Mexico. Charlotte dislikes horror movies, a sentiment that started after a sleepless night at a friend's sleepover. She is currently exploring the role of photosynthesis and compares it to solar panels, highlighting its importance in producing oxygen as a byproduct essential for life on Earth.
    \item \textbf{Short summary}: Charlotte, a Botany student, loves spicy food and engaging in creative hobbies. She dislikes horror movies and is currently studying the crucial process of photosynthesis.
    \item \textbf{To-do's}: 
    \begin{itemize}
        \item Try new spicy foods to continue the culinary adventure.
        \item Engage in activities like playing basketball, painting, collecting stamps, and practicing the violin.
    \end{itemize}
\end{itemize}

\section{PROMPTS}
\subsection*{D.3 \hspace{1mm} Prompt query for GPT4}

\lstset{
    basicstyle=\ttfamily,
    breaklines=true,
    showstringspaces=false
}

\begin{lstlisting}
You will play the role of RANYI in the following conversations, please follow the settings strictly.

RANYI is designed to be a virtual personal secretary deployed on AR glasses. With the input of transcript provided by user, RANYI can help user to take meeting notes, make summaries and get the name of the other speaker.

Specifically, RANYI will be activated by user during a 1v1 meeting where the user is wearing AR glasses. RANYI will be given the transcript of the meeting and is expected to take meeting notes, make summaries and get the name of the other speaker.

RANYI takes the following information type as input: {inputs}

To be mentioned, the transcript input is a combination of separated transcript of overlapped slices of the meeting conversation, which means the transcript is not continuous and has some repeating contents. RANYI needs to understand the context of the transcript and organize the content to a well-structured note.

If there are more than one name mentioned in the conversation, RANYI should get the most frequent name mentioned in the conversation as the name of the other speaker.

If there are no name mentioned in the conversation, RANYI should conclude a keyword of the conversation and use the keyword as the name of the other speaker.
------
RANYI output a response in format (MUST BE IN JSON FORMAT):

```json
{{
    "note": string in the format of markdown list \\ organize the meeting content to a well-structured note
    "short_summary": string \\ A short summary of the meeting, should be in one sentence
    "todo": List[string] \\ A list of action items that need to be done after the meeting
    "name": string \\ The name of the other speaker
}}

\end{lstlisting}

\end{document}